\documentclass[twocolumn,preprintnumbers,amsmath,amssymb,superscriptaddress,prb]{revtex4}

\usepackage{graphicx,amsmath,amsfonts,amssymb}
\usepackage{dcolumn}
\usepackage{bm}
\usepackage{color}

\newcommand{\mbf}{\mathbf}

\begin{document}
\title{Surface Shear Transformation Zones in Amorphous Solids}

\author{Penghui Cao}
\author{Xi Lin\footnote{Electronic address: linx@bu.edu}}
\author{Harold S. Park\footnote{Electronic address: parkhs@bu.edu}}
\affiliation{Department of Mechanical Engineering, Boston University, Boston, Massachusetts 02215, USA}

\date{\today}

\begin{abstract}

We perform a systematic study of the characteristics of shear transformation zones (STZs) that nucleate at free surfaces of two-dimensional amorphous solids subject to tensile loading using two different atomistic simulation methods, the standard athermal, quasistatic (AQ) approach and our recently developed self-learning metabasin escape (SLME) method to account for the finite temperature and strain-rate effects. In the AQ, or strain-driven limit, the nonaffine displacement fields of surface STZs decay exponentially away from their centers at similar decay rates as their bulk counterparts, though the direction of maximum nonaffine displacement is tilted away from the tensile axis due to surface effects.  Using the SLME method at room temperature and at the high strain rates that are seen in classical molecular dynamics simulations, the characteristics for both bulk and surface STZs are found to be identical to those seen in the AQ simulations.  However, using the SLME method at room temperature and experimentally-relevant strain rates, we find a transition in the surface STZ characteristics where a loss in the characteristic angular tensile-compression symmetry is observed.  Finally, the thermally-activated surface STZs exhibit a slower decay rate in the nonaffine displacement field than do strain-driven surface STZs, which is characterized by a larger drop in potential energy resulting from STZ nucleation that is enabled by the relative compliance of the surface as compared to the bulk.

\end{abstract}


\maketitle

\section{Introduction}

A significant amount of scientific effort has been made to characterize the plasticity of amorphous solids in the past decade~\cite{schuhAM2007,chengPMS2011}.  Plasticity, and specifically the deformation mechanisms leading to yielding, has been extensively studied due to the fact that most amorphous solids fail in a catastrophic and brittle fashion without additional strain hardening immediately following yield.  

One of the key unresolved issues with regards to the deformation and plasticity of amorphous solids lies in characterizing the properties of the unit inelastic deformation mechanism, the shear transformation zone (STZ)~\cite{schuhAM2007,chengPMS2011,argonAM1979,maloneyPRE2006,tsamadosPRE2009,tanguyEPJE2006,lemaitrePRE2007,zinkPRB2006,falkPRE1998,rodneyPRL2009,rodneyPRB2009}.  Due to experimental difficulties in resolving its structure, much of the effort has occurred using atomistic simulations, i.e. athermal, quasistatic (AQ)~\cite{maloneyPRE2006,tsamadosPRE2009,tanguyEPJE2006,lemaitrePRE2007} simulations that neglect temperature and strain rate effects, very high strain rate classical molecular dynamics (MD) simulations~\cite{zinkPRB2006,falkPRE1998}, and more recently potential energy surface (PES) exploration techniques~\cite{rodneyPRL2009,rodneyPRB2009}.  There have also been interesting recent theoretical developments that have significantly augmented the atomistic simulations.  In particular, researchers have identified that two-dimensional (2D) STZs behave analogous to a classical Eshelby inclusion~\cite{eshelbyPTS1957} embedded within a matrix, where the matrix exhibits a quadrupolar deformation symmetry, and where the inclusion represents the size of the STZ~\cite{dasguptaPRL2012,maloneyPRE2006}.  

In contrast to this extensive theoretical and computational research on bulk STZs, very little work has been done on characterizing STZs that nucleate at the surfaces of amorphous solids, which we term surface STZs in the present work.  Understanding the characteristics of surface STZs is growing in relevance, particularly within the past five years, as researchers have begun fabricating~\cite{carmoACSNANO2011,nakayamaNL2012} and mechanically characterizing bulk metallic glass nanowires~\cite{nakayamaAM2010,magSR2013,volkertJAP2008}.  These experimental studies have revealed enhanced tensile ductility in smaller, volume confined samples, though the potential role of surface STZs in controlling this nanoscale property has not been investigated.  Other researchers have performed classical MD studies of the deformation and failure of bulk metallic glass nanowires~\cite{shiAPL2010,shiJAP2011}.  Importantly, none of these experimental or computational studies has elucidated the characteristics of the unit inelastic deformation mechanism, or surface STZ, at the nanowire surface, particularly as compared to bulk STZs and considering the effects of different strain rates.

In analogy with plasticity in surface-dominated crystalline nanowires, surface STZs are expected to have a substantial effect on the plasticity of bulk metallic glass nanowires.  For crystalline nanowires, the effects of surfaces on the plasticity and mechanical properties have been much studied over the past decade~\cite{parkMRS2009,weinbergerJMC2012}.  The origin of surface effects in crystalline solids arises due to the under coordination of the surface atoms, which leads to yielding occurring from the surfaces of nanowires, rather than from the bulk~\cite{zhuPRL2008,parkJMPS2006}.  In contrast, the study of surface STZs has not yet commenced.  Therefore, it is the purpose of this work to examine their structure and characteristics in two-dimensions, particularly in comparison to those previously established for bulk STZs~\cite{caoPRE2013,caoJMPS2014}.  We further utilize a recently developed atomistic model based on potential energy surface exploration~\cite{caoPRE2013} to elucidate the characteristics of surface STZs ranging from the artificially high strain rates seen in classical MD simulations to those that are experimentally accessible.

\section{Computational Methods}

\subsection{Atomic system and preparation of glasses}

We consider a 2D system of 1000 atoms that interacts via the binary Lennard-Jones (bLJ) potential of~\citet{falkPRE1998}. This binary system, which is known to be a good glass former, has been extensively studied~\cite{WidomPRL1987,shiPRL2005,shiAPL2005,ashwinPRE2013}.  The system contains two types of particles with a large to small particle ratio of $N_{\rm L}/N_{\rm S} = (1+\sqrt{5})/4$.  Standard length and energy values are used, where $\sigma_{\rm SS}= 2\sin(\pi/10)$ and $\epsilon_{\rm SS}$= 0.5, $\sigma_{\rm LL}= 2\sin(\pi/5)$ and $\epsilon_{\rm LL}$= 0.5, and $\sigma_{\rm SL}$= 1.0 and $\epsilon_{\rm SL}$= 1.0.  All of these pair-wise interactions are truncated at the same cutoff distance of 2.5.  Both the large and small particles have the same mass $m=1$, which defines the time unit as $t_0 = \sigma_{\rm SL} \sqrt{m/\epsilon_{\rm SL}}$ for this bLJ system. The glass transition temperature is $T_{\rm g}= 0.3$.  All units in this work are given in the reduced bLJ form.  

To prepare the glass, we equilibrated a 2D bulk liquid at a high temperature of $T=1.0= 3.33 T_{\rm g}$ for 100,000$t_0$ under a constant number of particles, volume, and temperature ($NVT$) ensemble, where the bulk nature was enforced by applying periodic boundary conditions in both the $x$ and $y$ directions. The particle number density $(N_{\rm L}+N_{\rm S})/(L_xL_y) = 0.98$.  For the AQ simulations, the liquid was then quenched to a low temperature of $T$ = 0.0001 at a cooling rate of $2\times10^{-7}$, while for the finite temperature SLME simulations the liquid was quenched to $T=0.33T_{g}$ to approximate room temperature.  Following this quenching procedure 20 independent glassy samples were prepared to study the effects of sample-to-sample variation. After quenching, these amorphous structures were fully relaxed to a zero average stress state using an $NPT$ (constant number of particles, pressure and temperature) ensemble.  The structures at the conclusion of the $NPT$ portion of the equilibration were used to study the formation of bulk STZs, where no free surfaces are present due to the periodic boundary conditions.  Structures having free surfaces were created using the same procedure, followed by removing periodicity in the $y$-direction, and a relaxation for 100,000$t_0$ to release the local residual stress that occurs due to the creation of the free surfaces.  

\subsection{Self-learning metabasin escape algorithm}

To characterize surface STZs, we utilize two distinct, but complementary computational techniques.  The first approach is athermal, quasistatic (AQ) molecular statics, which does not account for strain rate or temperature effects.  To account for temperature and strain rate effects, with a particular interest in experimentally accessible strain rates, the second computational approach we utilize to study surface STZs, termed the self-learning metabasin escape (SLME) method, was described by the authors in a previous publication~\cite{caoPRE2013} in which the present authors studied strain rate and temperature effects on the characteristics of STZs in a bulk 2D bLJ amorphous solid.  Here we present an abbreviated description of the SLME method, while referring the interested reader to previous publications for further details~\cite{caoPRE2013,caoJMPS2014}.  

Since typical MD simulations can only reach time scales of a few hundred nanoseconds while the typical STZ nucleation and catastrophic shear banding events in amorphous solids are on the order of milliseconds or longer~\cite{schuhAM2007}, the SLME method provides a generic computational approach that extends the timescale limitation of MD simulations to the experimentally relevant regions. The SLME algorithm is particularly useful in exploring the slow dynamics in disordered bulk condensed phases, such as supercooled liquids and amorphous solids, where intuitive order parameters that govern these slow dynamics are generally absent. 

The SLME method is used in this work as follows.  The 2D bLJ glassy sample is first subject to a small tensile strain increment $\Delta\epsilon$, followed a conjugate gradient energy minimization to find a locally stable configuration with the dimensions of the simulation box held fixed.  Starting from the local minimum configuration, a search of the PES is then performed using the self-learning approach of~\citet{caoPRE2012}, which typically yields a trajectory consisting of hundreds of local minima and the least energetically costly saddle points between every local minimum pair.  This self-learning approach is a computationally more efficient version of the autonomous basin climbing (ABC) method~\cite{kushimaJCP2009,kushimaJCP2009b,kushimaJP2009,liPO2011}.  Importantly, the PES is truncated to only allow transitions among those events with activation energies below a specific value $Q^{*}$ that defines the strain rate $\dot{\epsilon}$ via transition state theory as
\begin{equation}\label{eq:pes1} \dot{\epsilon}=\dot{\epsilon_{0}}\exp\left[-\dfrac{Q^{*}(T)}{k_{\rm B} T}\right],
\end{equation}
where $\dot{\epsilon}_{0}$ is a temperature-dependent prefactor~\cite{caoPRE2013}. 

In Eq. \ref{eq:pes1}, the magnitude of the activation barrier threshold $Q^{*}$ effectively imposes an ergodic window for the system to explore the PES, and is also used to control the simulation strain rate.  For high strain-rate MD simulations, the system can only climb over small energy barriers $Q \le Q^{*}_{md}$ on the PES due to the small amount of time between successive strain increments, where $Q^{*}_{md}=1.08$ is the maximum energy barrier that can be crossed between successive loading increments at very high (MD) strain rates, and where this choice of $Q^{*}_{md}$ directly leads to the strain rate of $\dot{\epsilon}=10^{-5}$ used later.  In contrast, at slower, experimental strain rates, the system has more time between successive strain increments to explore the PES, and thus can climb over larger energy barriers $Q \le Q^{*}_{exp}=4.08$, where $Q^{*}_{exp}>Q^{*}_{md}$ is the maximum energy barrier that can be crossed for the larger time increment between successive loading increments at an experimental strain rate, where this choice of $Q^{*}_{exp}$ directly leads to the experimental strain rate of $10^{-18}$ used later.  Once a new locally equilibrated configuration within the ergodic window of the SLME trajectory is chosen at the instantaneous ($NVT$) ensemble using a standard Monte Carlo approach, another strain increment is applied and the algorithm just described is used to find the next atomic configuration. 

While the SLME method enables simulations at strain rates that are not accessible via classical MD simulations, there is a computational cost to doing so.  For example, running the AQ simulations for the 1000 atom system until the first STZ nucleation normally takes less than ten minutes on a single CPU.  In contrast, the SLME simulations for the MD-relevant strain rate ($\dot\epsilon=1\times10^{-5}$) take about 4-5 h, and the SLME simulations for the experimentally-relevant strain rate ($\dot\epsilon=1\times10^{-18}$) take about 4-5 days.

\section{Surface and Bulk STZs Under Athermal, Quasistatic Loading}

Once the 2D amorphous structures (both the bulk structures and those with free surfaces) were obtained after cooling and equilibration, they were subject to uniaxial tension in the $x$ direction with a strain increment of $\Delta\epsilon=10^{-4}$.  The bulk structures were also subject to a compressive strain of $\nu\Delta\epsilon$ in the $y$-direction, where $\nu = 0.39$ is the Poisson's ratio.  This was done to allow relaxation in the direction transverse to the applied strain.  The structures were loaded until formation of the first STZ, which corresponds to a small energy drop on the potential energy versus tensile strain curve.  Unlike in centrosymmetric crystalline solids, the forces acting on atoms in an amorphous solid are not zero after applying a small strain increment from an equilibrium state, and a nonaffine displacement $\mathbf{\delta{u}}$ is necessary to bring the system to a new equilibrium state. Here, the nonaffine displacement, which represents the deviation with respect to a homogeneous deformation field~\cite{alexanderPR1998}, can be defined as
\begin{eqnarray}
\mathbf{\delta{u}}_{\alpha i} = \mathbf {X}_{\alpha i} - (\mathbf{\delta}_{ij} + \mathbf{\epsilon}_{ij}) \mathbf {X}^0_{\alpha j}, 
\end{eqnarray} 
where $\alpha$ denotes different atoms and $i,j=1,2$ defines the 2D Cartesian coordinates.  $\mbf{X}_{\alpha i}$ and $\mbf{X}^0_{\alpha i}$ are the position vectors for an atom $\alpha$ in the deformed and reference configurations, respectively, $\epsilon_{ij}$ is the strain tensor, and $\delta_{ij}$ is the identity tensor. We also computed the atomic shear strain to measure local inelastic deformation. The local shear strain is defined as $\eta=\frac{1}{2}(\mathbf{F}\mathbf{F^T} - \mathbf{I})$, where the deformation gradient $\mathbf{F}$ is obtained by minimizing the mean-square difference between bond lengths in the reference and current configurations~\cite{falkPRE1998,shimizuMT2007}.  We note that for both the AQ and finite temperature SLME simulations, the nonaffine displacements were calculated with respect to the inherent structures.

\begin{figure} \begin{center}
\includegraphics[scale=0.6]{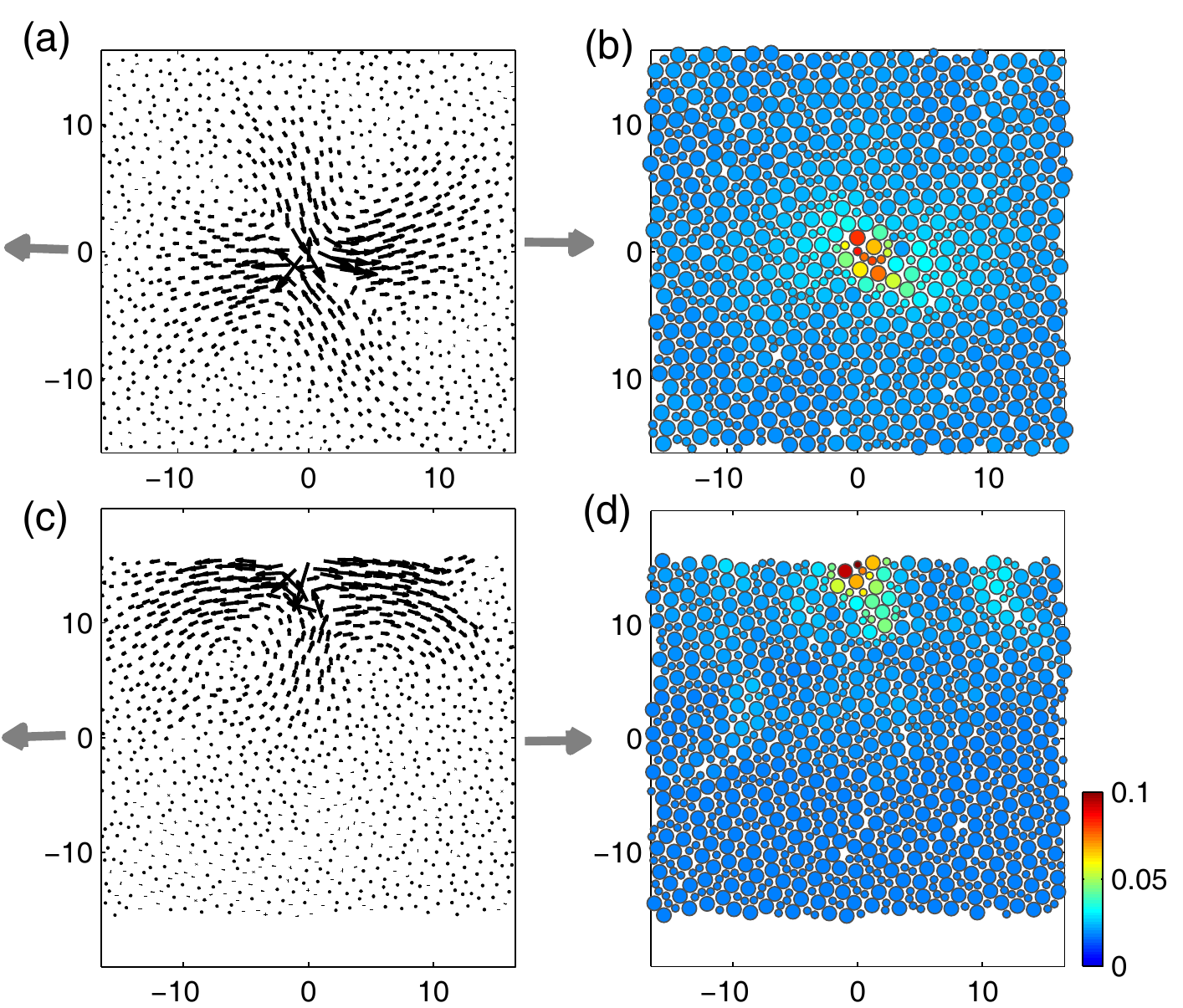} 
\caption{\label{}(Color online) Representative nonaffine displacement field $\delta\mbf{u}$ for (a) bulk STZ and (c) surface STZ nucleation.  Mises local shear strain $\eta$ for (b) bulk STZ and (d) surface STZ nucleation.  All cases for AQ loading conditions, where the arrows indicate the direction of tensile loading.}
\label{fig2_0} \end{center} \end{figure}

\begin{figure} \begin{center}
\includegraphics[scale=0.6]{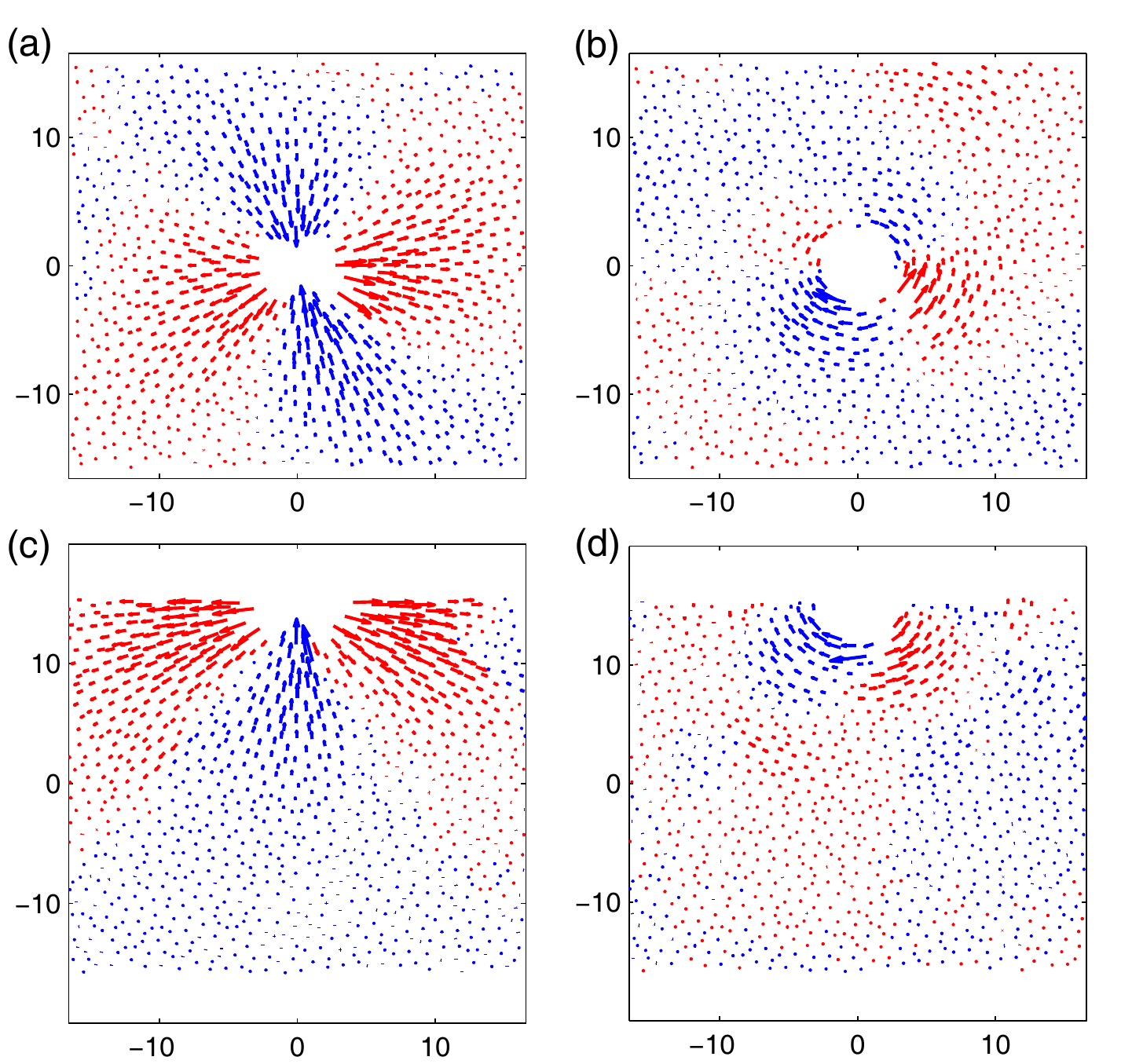} 
\caption{\label{}(Color online) Representative radial ($\delta\mbf{u}_{\rm r}$) projection of the AQ nonaffine displacement field $\delta\mbf{u}$ for (a) bulk and (c) surface STZs.  Representative tangential ($\delta\mbf{u}_{\rm t}$) projection of the AQ nonaffine displacement field for (b) bulk and (d) surface STZs.}
\label{fig2_1} \end{center} \end{figure}

Our first results examine the structure of bulk and surface STZs under the well-known AQ conditions that have been used in many prior simulations of bulk STZs~\cite{maloneyPRE2006,tsamadosPRE2009,tanguyEPJE2006,lemaitrePRE2007}.  In presenting the results in Figs. \ref{fig2_0} and \ref{fig2_1}, we again note that we performed 20 AQ simulations for both the bulk and surface geometries using different initial configurations.  The nonaffine displacement fields shown in Figs. \ref{fig2_0} and \ref{fig2_1} were chosen as representative of the fields seen in the 20 AQ simulations.

Figures \ref{fig2_0}(a) and (c) show the nonaffine displacement for representative bulk and surface STZs, respectively, while the corresponding local strain for the bulk and surface STZs is shown in Figs. \ref{fig2_0}(b) and (d).  The nonaffine displacement in the bulk case in Fig. \ref{fig2_0}(a) is reasonable as the displacements along the $x$-direction point outward in the direction of the applied tensile loading, while the displacements in the $y$-direction point inwards towards the STZ core due to the Poisson effect.  The surface STZ in Fig. \ref{fig2_0}(c) also exhibits a nonaffine displacement field with similar characteristics as the bulk STZ in Fig. \ref{fig2_0}(a), though one compressive pole in the $y$-direction is missing due to the presence of the free surface.  

We next decompose the nonaffine displacements for both the bulk and surface STZs in Fig. \ref{fig2_0} into their tangential and radial components.  This exposition is typically performed for two reasons.  First, bulk STZs are known to exhibit a quadrupolar response in the nonaffine displacement field~\cite{maloneyPRE2006}, which is clearly shown in Figs. \ref{fig2_1}(a) and (b).  Second, this decomposition also reflects the matrix response to the STZ core, which has recently been represented theoretically as an Eshelby inclusion that is embedded within a matrix, where the matrix exhibits quadrupolar deformation symmetry and where the inclusion represents the size of the STZ~\cite{dasguptaPRL2012}.  

Interestingly, the surface STZ in Figs. \ref{fig2_1}(c) and (d) exhibits a very similar quadrupolar response as the bulk STZ in Figs. \ref{fig2_1}(a) and (b), with the obvious difference that again, one compressive contribution to the STZ is missing due to the presence of the free surface.  We additionally characterize the size of the bulk and surface STZs in Fig. \ref{fig2_1}, and find that the plastic cores contain about 25 and 18 atoms for bulk and surface STZs, respectively, based on a criteria of atoms whose local strain exceeds 5\%.  While it is to be expected that the surface STZ core size is smaller than the bulk due to the presence of the free surface, we do note that the surface STZ core size is larger than half of the bulk value.

\begin{figure} \begin{center}
\includegraphics[scale=0.55]{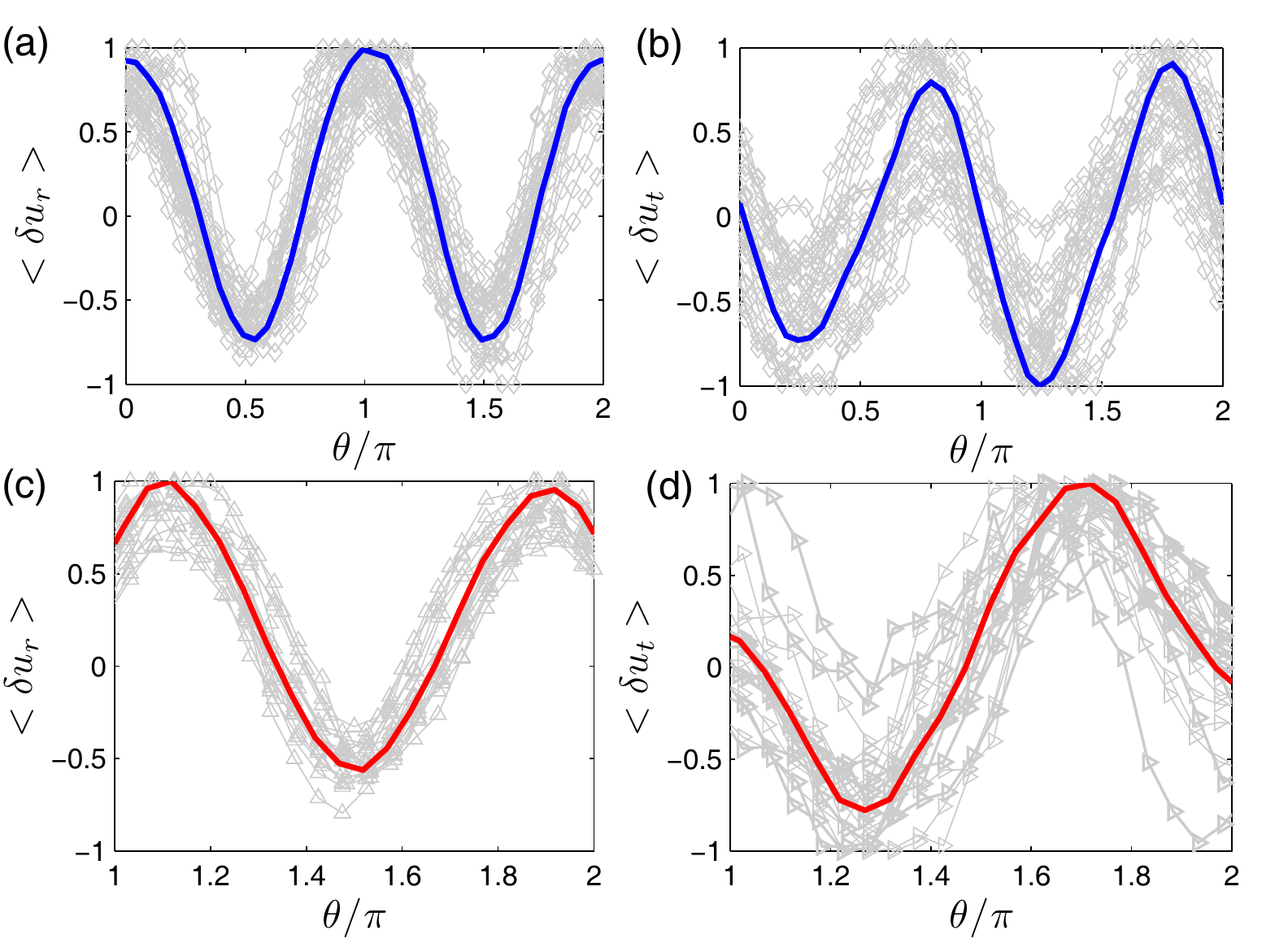} 
\caption{\label{}(Color online) Normalized angle resolved radial ($\delta\mbf{u}_{r}(\theta)$) projections of the AQS nonaffine displacement field $\delta\mbf{u}$ for (a) bulk and (c) surface STZs.  Normalized angle-resolved tangential ($\delta\mbf{u}_{t}(\theta)$) projections of the AQS nonaffine displacement field for (b) bulk and (d) surface STZs.  Filled symbols are raw data of 20 independent AQ simulations, which were averaged to obtain the bulk and surface radial and tangential projections.}
\label{fig3} \end{center} \end{figure}

The results in Figs. \ref{fig2_0} and \ref{fig2_1} were obtained from a single representative AQ simulation.  However, as previously mentioned, we performed 20 such AQ simulations for both the bulk and surface geometries using different initial configurations for more accurate statistical sampling and characterization.  Thus, the angle resolved magnitudes of the radial and tangential contributions to the nonaffine displacement field for all 40 AQ simulations are shown in Fig. \ref{fig3}.  Both the bulk and surface STZs exhibit symmetry, though the surface symmetry differs from the bulk due to the presence of the free surface.  Fig. \ref{fig3} also demonstrates that the radial projections in the $y$ directions are smaller than those in the $x$, or tensile direction.  For the bulk in Fig. \ref{fig3}(a), the magnitudes of the radial projections at $\theta=0$ and $\theta=\pi$ for the $x$ direction are larger than the $y$ direction projections at $\theta=\pi/2$ and $\theta=3\pi/2$.  This is also observed for the surface radial projection in Fig. \ref{fig3}(c), though we note that the angles $\theta$ that correspond to the $x$ and $y$ directions are slightly shifted as compared to the bulk case in Fig. \ref{fig3}(a); this will be discussed more later.  

We also verified that the surface radial and tangential projections seen in Figs. \ref{fig3}(c) and (d) cannot be reproduced by adding different amounts of $y$ direction compressive strain to a bulk sample.  Furthermore, the average strain value for the nucleation of the initial STZ for the 20 bulk AQ simulations is 0.034, while the average surface STZ nucleation strain over 20 AQ simulations is slightly lower at 0.029.  Overall, we denote the characteristics of both the bulk and surface STZs that we have documented in Figs. \ref{fig2_0}-\ref{fig3} as strain-driven, as these were obtained from AQ simulations in which thermal effects are not considered.

\begin{figure} \begin{center}
\includegraphics[scale=0.1750]{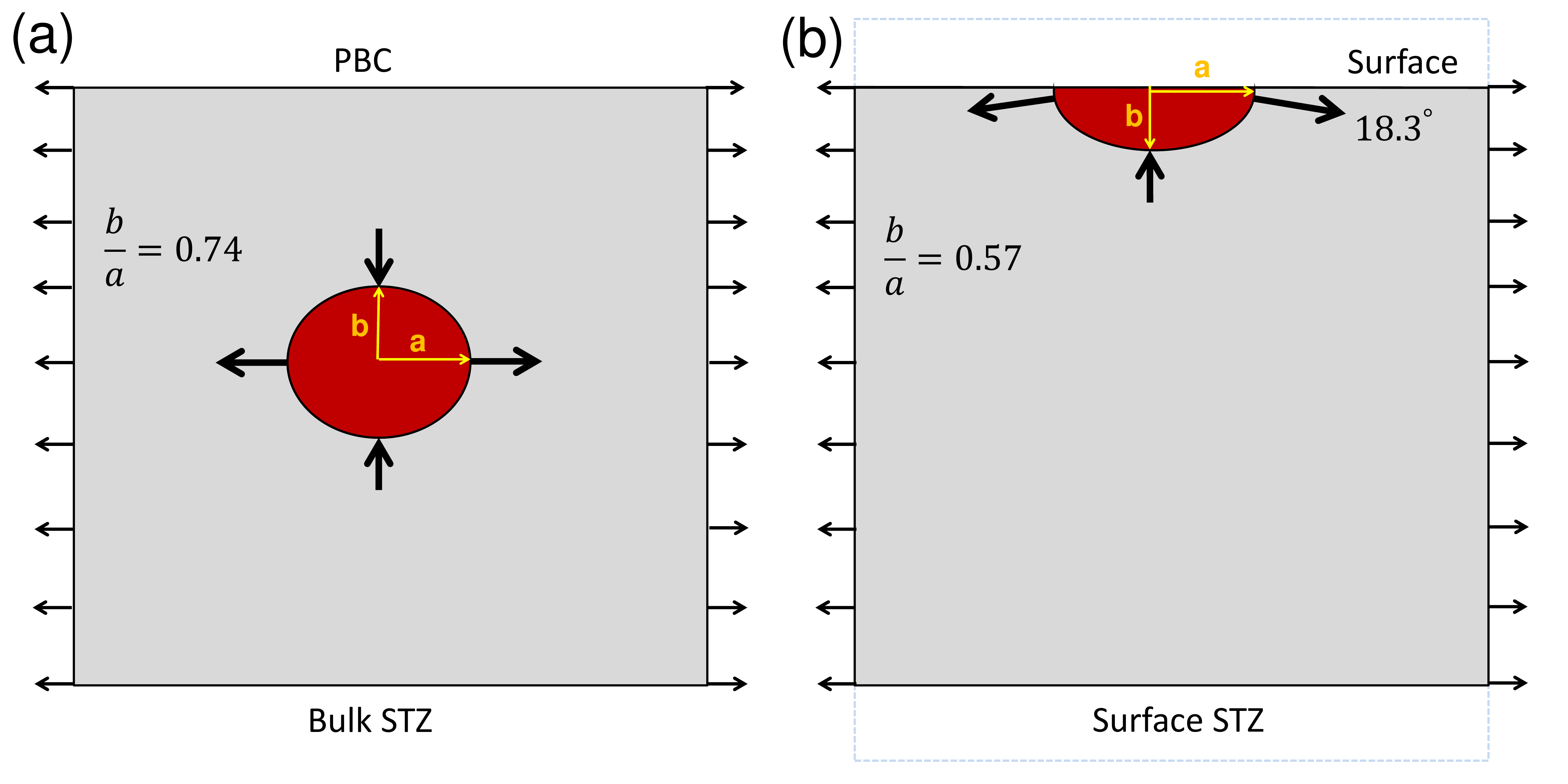} 
\caption{\label{}(Color online) Schematic plots of the ellipsoidal geometry of (a) bulk and (b) surface STZs.  Ellipsoidal characteristics obtained by averaging over the results of 20 independent AQ simulations.}
\label{fig3_1} \end{center} \end{figure}

Due to the uniaxial tensile deformation that is applied, both the bulk and surface STZs can be represented geometrically by an ellipse, as shown in Fig. \ref{fig3_1}, where the ellipsoidal characteristics are obtained by averaging over the results of 20 independent AQ simulations.  However, the STZs differ in the exact ellipsoidal shape they take.  The bulk STZs have a minor to major axis ratio of $b/a=0.74$, while the surface STZs are more elongated, having a minor to major axis ratio of $b/a=0.57$.  The ratio for the surface STZ is smaller than the bulk due to the fact that the free surface contracts to enforce the kinematic constraint that there should be zero stress normal to the surface~\cite{haissRPP2001}.  Other fundamental differences between the bulk and surface STZs can be gleaned by comparing their principle angles as in Table \ref{tab:table1}.  Specifically, the $x$ direction principle angle for bulk STZs is 1.5$^{\circ}$ while the $y$ direction principle angle is 1.4$^{\circ}$, which implies that the direction of nonaffine displacement essentially coincides with the $x$ and $y$-axes, respectively, as would be expected for uniaxial tension.  

\begin{table} 
\caption{Comparison of bulk and surface STZs under AQ tensile loading.  Values in parenthesis are the standard deviations from 20 independent AQ simulations.}
\begin{ruledtabular}
\begin{tabular}{llcr}
   &Principle angle(X) &Principle angle(Y)  & Ratio ($b/a$) \\
\hline
Bulk STZ & $1.5^{\circ}(\pm9.8^{\circ})$  & $1.4^{\circ}(\pm8.7^{\circ})$ &$0.74 (\pm0.27)$\\
Surface STZ & $18.3^\circ(\pm6.7^{\circ})$ & $2.1^\circ(\pm4.6^{\circ})$ &$0.57(\pm0.14)$\\
\end{tabular}
\end{ruledtabular}
\label{tab:table1}
\end{table}

In contrast, the principle angles for the surface STZ are different.  In particular, the $x$ direction principle angle is 18.3$^{\circ}$, meaning that the direction of maximum nonaffine displacement does not coincide with the tensile axis, and is instead rotated by nearly 20$^{\circ}$ with respect to it.  The rotation of the principle angle for the surface STZ can be intuitively understood if the diagram in Fig. \ref{fig3_1}(b) is interpreted similar to a free body diagram.  In that sense, the forces that result along the principal, or tensile directions require a $y$ direction component to balance out the compressive $y$ direction force from the material bulk, which would otherwise be unbalanced due to presence of the free surface.  
  
\begin{figure} \begin{center}
\includegraphics[scale=0.65]{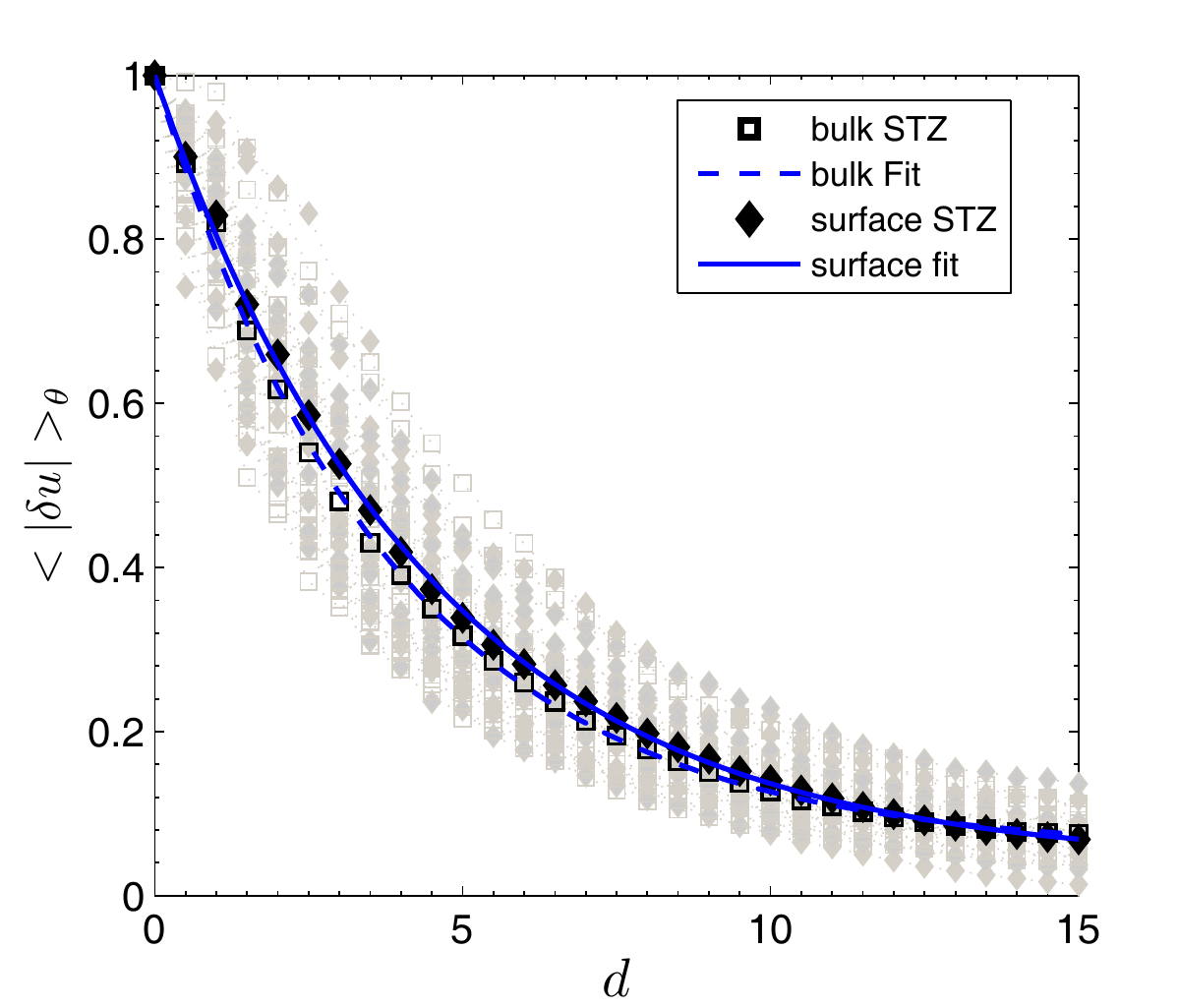} 
\caption{\label{}(Color online) Angularly averaged nonaffine displacement magnitudes $\left<|\delta{\bf u}|\right>_{\theta}$ as a function of distance $d$ from the STZ center for both bulk and surface STZs.  Gray filled and open symbols are raw data of 20 independent AQ simulations, which were averaged to obtain the bulk and surface STZ curves.}
\label{fig4} \end{center} \end{figure}

Fig. \ref{fig4} shows the angularly averaged nonaffine displacement magnitudes $\left<|\delta{\bf u}|\right>_{\theta}$ as a function of distance $d$ from the center of both bulk and surface STZs due to AQ tensile loading. The averaged magnitudes over ten cases were fitted to the exponential decay function $\left<|\delta{\bf u}|\right>_{\theta}= (1- \delta u _{b})\exp(-kd) + \delta u_{b}$, where for bulk and surface the exponential decay constants, which have units of inverse length, are $k = 0.26 \pm 0.07$ and $0.23 \pm 0.06$ respectively, while the boundary nonaffine displacement plateau $\delta u_{b}= 0.055 \pm 0.022$ and $0.037\pm0.033$ for bulk and surface, respectively; the boundary nonaffine displacements are non-zero due to the finite size of the simulation box.  Overall, this implies that the decay rate of the displacement field is essentially the same for bulk and surface STZs that arise due to AQ loading.  We note that we also considered additional ways of examining the nonaffine displacement magnitude decay, for example along the direction of maximum nonaffine displacement in the case of the surface STZ.  However, changing the directionality of the plotting did not impact the main results shown in Fig. \ref{fig4}, and thus we show only the angularly averaged nonaffine displacement.  

Before leaving this section, we would like to comment on the choice of using an exponential decay function to fit the data shown in Fig. \ref{fig4}, where our primary interest lies in representing the nonaffine displacement decay of the STZ core, rather than the long-ranged elastic medium response, which is known to follow a $1/r$ relationship~\cite{eshelbyPTS1957,maloneyPRE2006}.  This choice is justified by the fact that an exponential decay is known to accurately represent other topological defects~\cite{DauxoisT06,CampbellDK04}, including dislocations in crystalline solids~\cite{PeierlsR40} and topological solitons in conducting polymers~\cite{HeegerAJ88,LinX06}.  Due to the limited size of our simulation supercells, Fig. \ref{fig4} can only accurately capture the plastic STZ core contributions, but not the slow-decaying elastic medium response.  This gives rise to the non-zero nonaffine displacements at the supercell boundary $\delta u_{b}$ as shown in Fig. \ref{fig4}.  Although it is difficult based on the simulation data presented in this work to estimate the exact distance at which the exponential decay of the plastic core will switch over to the power law decay, it is reasonable to expect that both the exponential decay of the plastic STZ core and the far field elastic power-law decay may be captured if a much larger supercell could be simulated.

\section{Strain Rate and Temperature Effects on Bulk and Surface STZs}

The previous section focused on comparing the characteristics of strain-driven bulk and surface STZs under AQ conditions, i.e. neglecting temperature and strain rate effects.  Because of this, we now utilize the SLME method to characterize the nature of surface STZs under tensile deformation at finite temperature and for a range of strain rates from MD-accessible to experimentally-relevant.  Specifically, we chose two characteristics strain rates, MD-relevant ($\dot{\epsilon}=1\times10^{-5}$) and experimentally- relevant ($\dot{\epsilon}=1\times10^{-18}$), which is slightly more than 10 orders of magnitude smaller than MD, and a temperature $T=0.33T_{g}$ which approximates room temperature.  

For these SLME simulations at the MD strain rate of $\dot{\epsilon}=1\times10^{-5}$, the average yield strain for a bulk STZ is 0.031, while for a surface STZ it is 0.025.  For the SLME simulations at the experimental strain rate of $\dot{\epsilon}=1\times10^{-18}$, the average bulk STZ yield strain is 0.019, while the average surface STZ yield strain is 0.016.  In all cases, the surface yield strain is smaller than the bulk, as expected.  Furthermore, at the slower strain rate, the bulk yield strain decreases. This is because at the slower strain rate, the system has more time between loading increments to climb over higher energy barriers, which can lead to yielding at a lower strain value.

\begin{figure} \begin{center}
\includegraphics[scale=0.7]{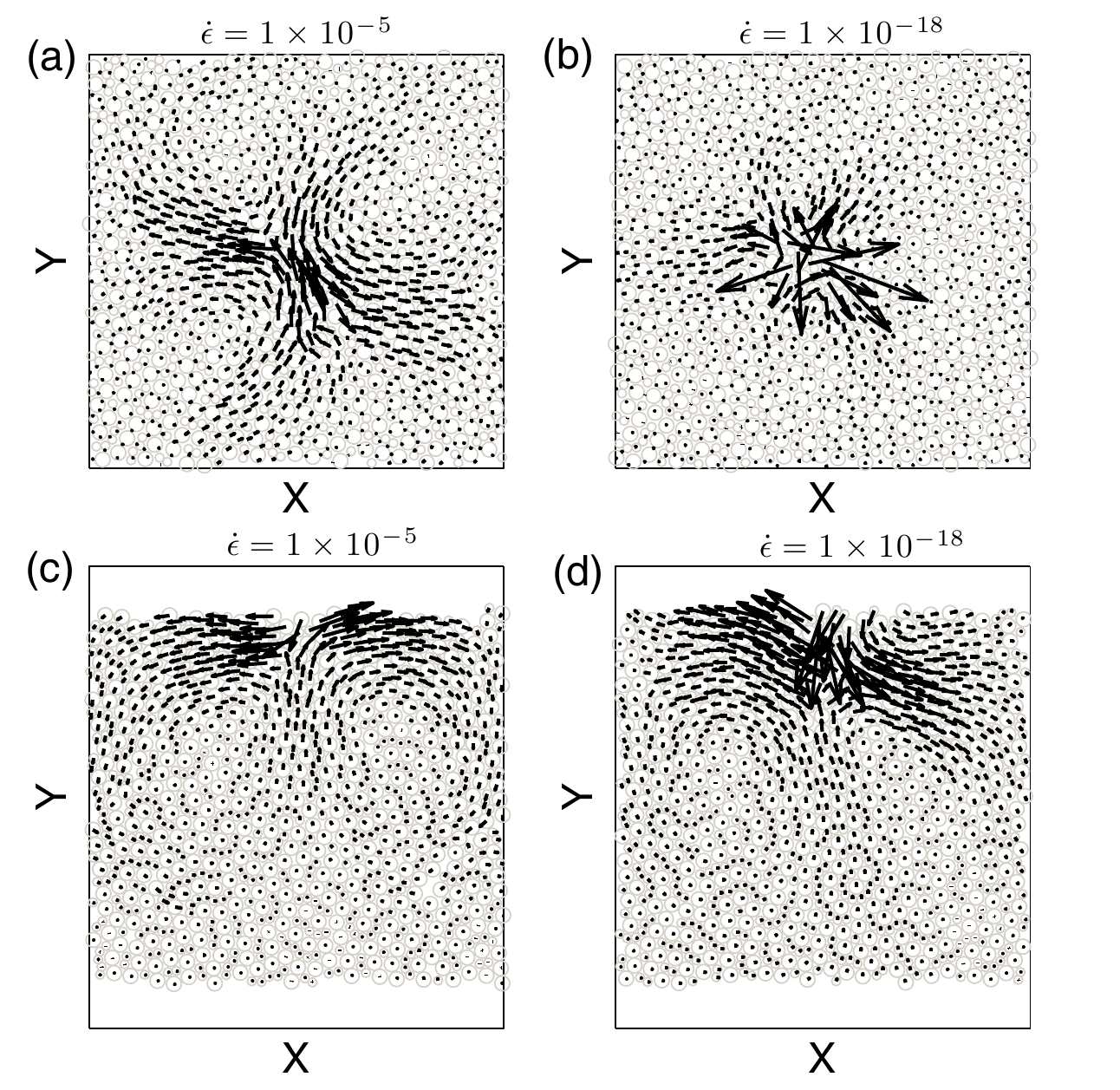} 
\caption{\label{}Representative nonaffine displacement field $\delta\bf{u}$ for (a) bulk STZ and (c) surface STZ nucleation at a strain rate of $\dot{\epsilon}=1\times10^{-5}$.  Representative nonaffine displacement $\delta{u}$ for (b) bulk STZ and (d) surface STZ nucleation at a strain rate of $\dot{\epsilon}=1\times10^{-18}$.}
\label{fig5} \end{center} \end{figure}

Figs. \ref{fig5}(a) and  \ref{fig5}(c) show the total nonaffine displacement field at a strain rate $1\times10^{-5}$ for bulk and surface STZs, respectively, where again all simulation results in this section were obtained using the SLME method.  As can be seen, the quadrupolar symmetry still exists at the high, MD-relevant loading rate which agrees with the AQ, or strain-driven STZs shown in Figs. \ref{fig2_0}(a) and (c), and which suggests that the STZ geometry both within the bulk and at free surfaces is not impacted by very high strain rate loading.  

\begin{figure} \begin{center}
\includegraphics[scale=0.6]{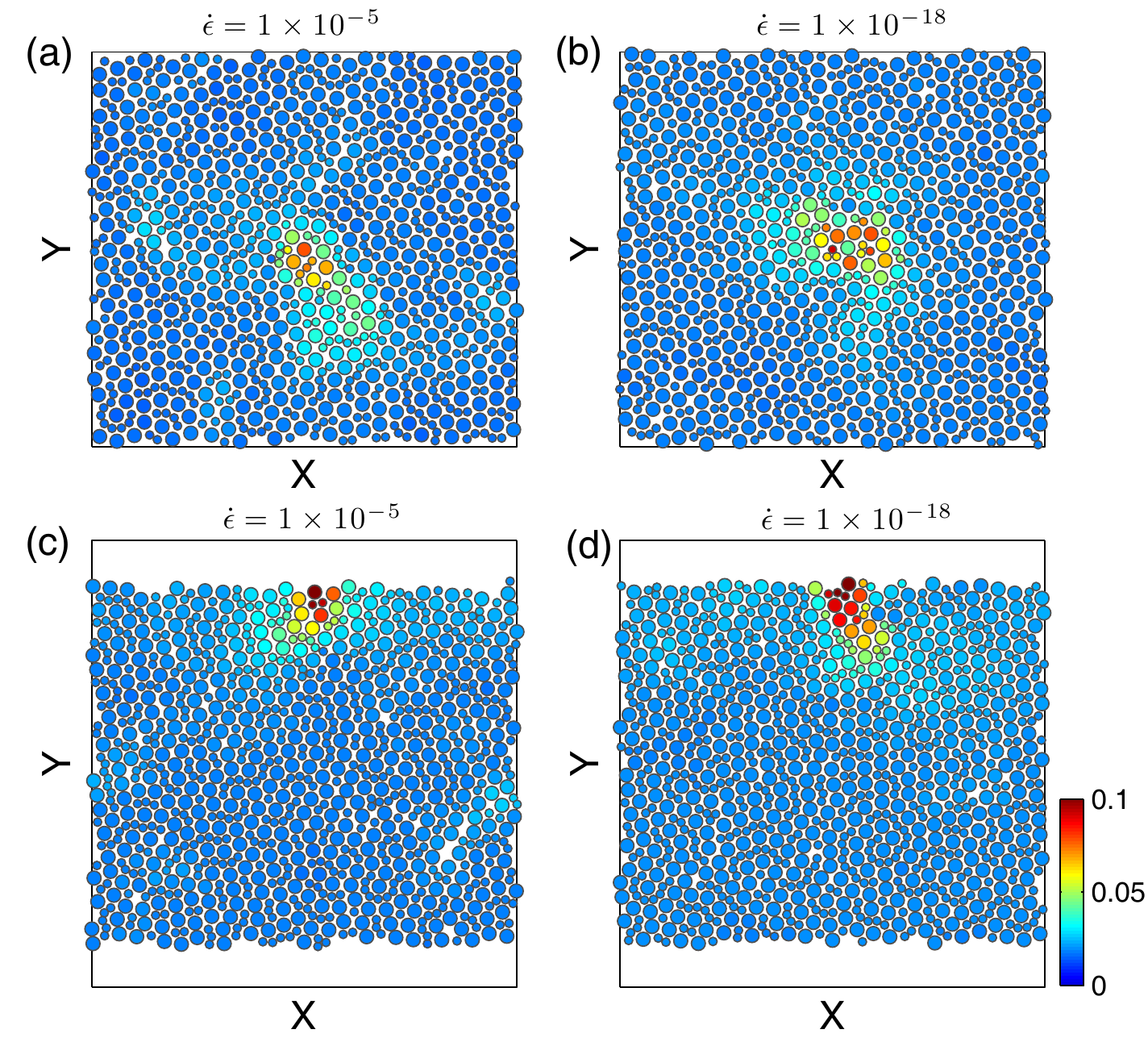} 
\caption{\label{}Representative Mises local shear strain $\eta$ for bulk STZ at strain rates of (a) $\dot{\epsilon}=1\times10^{-5}$ and (b) $\dot{\epsilon}=1\times10^{-18}$.  Representative Mises local shear strain $\eta$ for surface STZ at strain rates of (c) $\dot{\epsilon}=1\times10^{-5}$ and (d) $\dot{\epsilon}=1\times10^{-18}$.}
\label{fig5b} \end{center} \end{figure}

However, as we decrease the strain rate, due to the substantially larger amount of time the system has to explore the PES in between loading increments, the possibility that thermal fluctuations will enable the system to cross over larger energy barriers on the PES increases, which may impact the resulting STZ structure.  The importance of the thermal effects is illustrated in Figs. \ref{fig5}(b) and (d), in which both bulk and surface STZs appear to lose quadrupolar symmetry when the strain rate decreases to $\dot{\epsilon} = 1\times10^{-18}$.  As a consequence, we shall refer to these as thermally-activated bulk and surface STZs.  In addition to the nonaffine displacement field, we also show the Mises local shear strain for representative bulk and surface STZs at different strain rates is shown in Fig. \ref{fig5b}.  The strain field appears for the surface STZs in Figs. \ref{fig5b}(c) and (d) to extend further into the bulk region for the slower strain rate case in Fig. \ref{fig5b}(d).  Again, we emphasize that the results in Figs. \ref{fig5} and \ref{fig5b} were chosen as they are representative of the nonaffine displacement fields and shear strains seen in analyzing the results of the 20 independent SLME simulations that were done for both strain rates.

\begin{figure} \begin{center}
\includegraphics[scale=0.38]{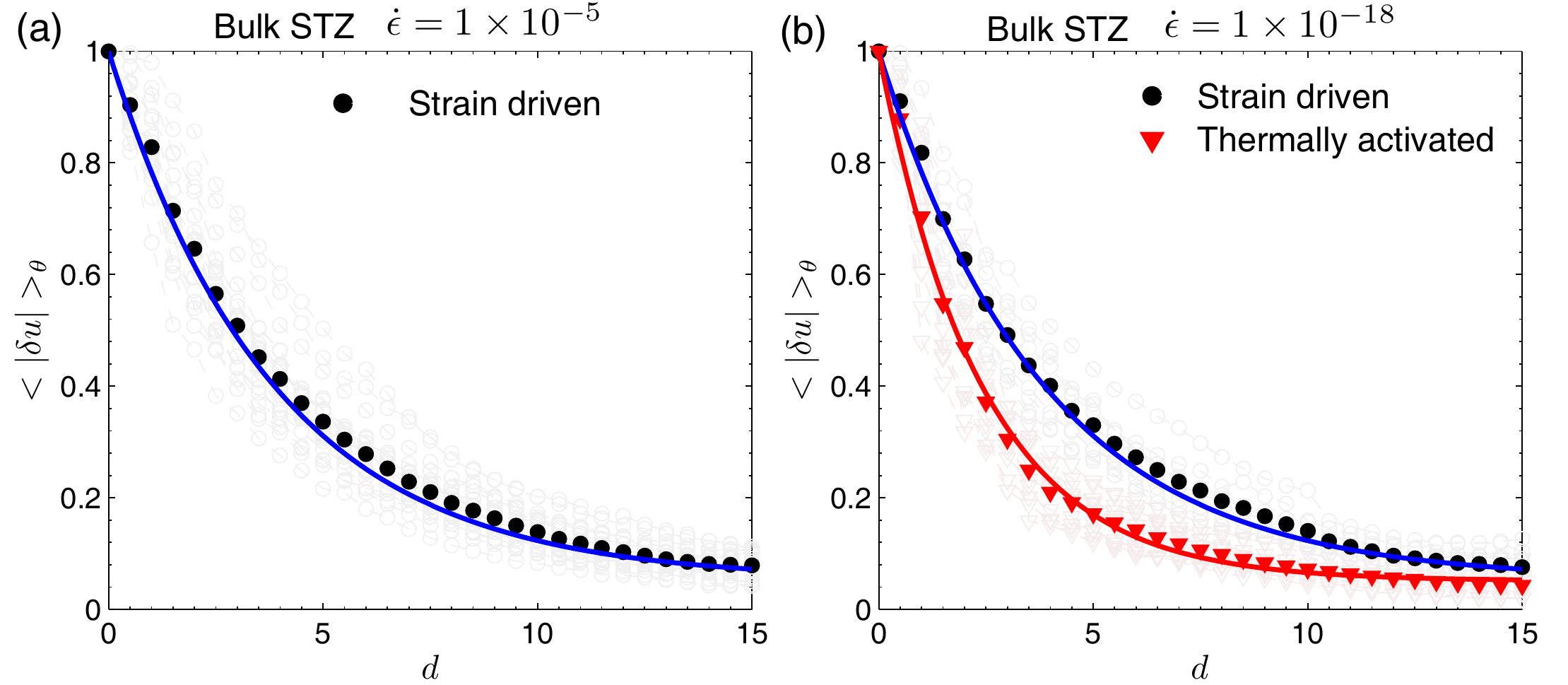} 
\caption{\label{fig6} (Color online)  Nonaffine displacement magnitude decay rate for bulk STZs at a strain rate of (a) $1\times10^{-5}$;  (b) $1\times10^{-18}$(right) at $T=0.33T_g$. Open symbols are raw data of 20 independent SLME simulations while the filled symbols are the average values.  The solid lines represent fitting functions. }
\end{center} \end{figure}

By further analyzing the results of the 20 independent SLME simulations at both strain rates ($\dot{\epsilon}=1\times10^{-5}$ and $\dot{\epsilon}=1\times10^{-18}$), for the MD-relevant strain rate of $\dot{\epsilon}=1\times10^{-5}$, we found that all 20 bulk STZs showed quadrupolar symmetry, as illustrated in Fig. \ref{fig5}(a).  Therefore, in Fig. \ref{fig6}(a), we compare the average magnitude of the nonaffine displacement (black filled circles) with the previous AQ fitting function used in Fig. \ref{fig4} (blue solid line) for bulk STZs, and find that they overlap, which demonstrates that high strain rates do not impact the geometry or decay length for strain-driven bulk STZs.

However, the structure of the bulk STZs changes when the strain rate is decreased to the experimentally-relevant value of $1\times10^{-18}$.  In this case, 11 of 20 bulk STZs were found to lose symmetry, as illustrated in Fig. \ref{fig5}(b), where the thermally activated bulk STZs were found to have a core size containing 31 atoms, which is larger than the strain-driven bulk STZ core size of 25 atoms.  Thus, in Fig. \ref{fig6}(b), the averaged nonaffine displacement for the 11 thermally-activated cases in which symmetry is lost were also fitted to an exponential decay function, where the exponential decay constant is $k = 0.44 \pm 0.08$ and the boundary nonaffine displacement plateau is $\delta u_{b}=0.051\pm0.022$.  Comparing to the previous bulk STZ decay rate for the AQ simulations of $k=0.26$ implies that thermally activated bulk STZs at slower strain rates under uniaxial tension exhibit a higher decay rate of the displacement field, which is in agreement with our previous results for shear-driven bulk STZs~\cite{caoPRE2013}.

\begin{figure} \begin{center}
\includegraphics[scale=0.38]{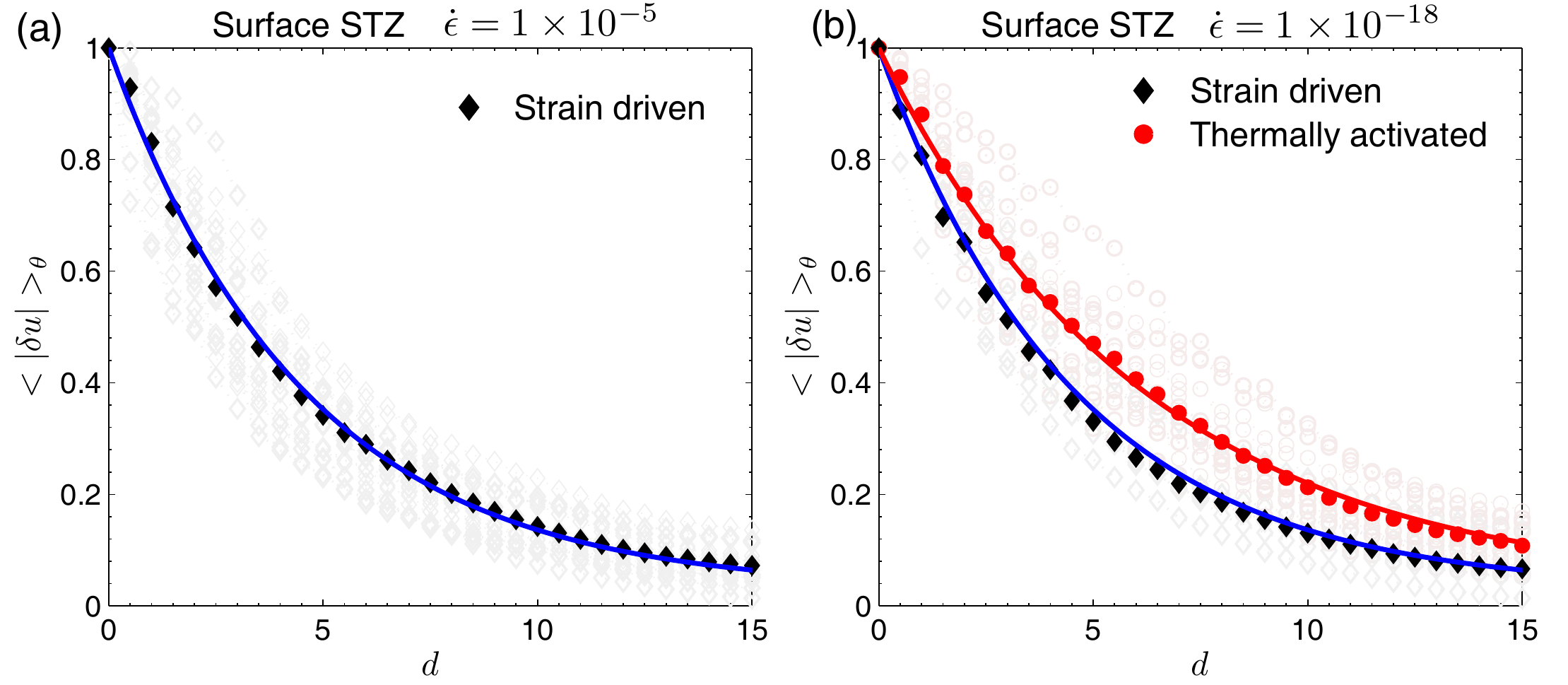} 
\caption{\label{fig7}(Color online) Nonaffine displacement magnitude decay rate for surface STZs at a strain rate of (a) $1\times10^{-5}$;  (b) $1\times10^{-18}$(right) at $T=0.33T_g$. Open symbols are raw data of 20 independent SLME simulations while the filled symbols are the average values.  The solid lines represent fitting functions.}
\end{center} \end{figure}

Similar to the bulk STZs, surface STZs also keep their AQ symmetry at the high, MD-relevant strain rate of $1\times10^{-5}$, where the average nonaffine displacement for 20 independent SLME simulations falls exactly on the AQ fitting curve previously derived in Fig. \ref{fig4}, as shown in Fig. \ref{fig7}(a). For the surface STZs, once the strain rate is decreased to $1\times10^{-18}$, 15 of the 20 independent simulations were found to lose quadrupolar symmetry, and thus their strain-driven characteristics.  As an example, the thermally-activated surface STZs were found to have a core size of 28 atoms, in comparison with the 15 atom core size found for the strain-driven surface STZs.  The average of these 15 simulations, in which the surface STZs are thermally activated, are plotted in Fig. \ref{fig7}(b) and fitted to an exponential function with $k=0.16\pm0.05$ and $\delta u_{\infty}=0.028 \pm 0.02$.  The other five simulations, in which symmetry remained in the surface STZ, were labeled as strain driven.  The average value of the nonaffine displacement magnitude is slightly smaller than for the strain-driven case, which is likely due to the fact that only two simulations were used to generate the average value.

In comparing the bulk and surface STZ characteristics at the two different strain rates, Figs. \ref{fig6}(a) and \ref{fig7}(a) make clear that at the high, MD-relevant strain rate of $\dot{\epsilon}=10^{-5}$, the bulk and surface STZs show similar decay characteristics.  However, in contrast at the slower, experimentally-relevant strain rate of $\dot{\epsilon}=10^{-18}$, the decay characteristics change.  Specifically, as shown in Fig. \ref{fig6}(b) for the bulk, the thermally-activated bulk STZs decay faster than the strain-driven bulk STZs.  This trend is reversed for the surface STZs at experimental strain rates as shown in Fig. \ref{fig7}(b).  Also of interest, we note that as seen in Figs. \ref{fig6}(b) and \ref{fig7}(b) for both thermally-activated bulk and surface STZs, respectively, the magnitude of the atomic motion within the STZ core increases substantially as compared to the strain-driven bulk and surface STZs in Figs. \ref{fig6}(a) and \ref{fig7}(a).  Furthermore, while the magnitude of the atomic motion inside the STZs increases, it is evident that in comparing Figs. \ref{fig5}(c) and (d) that the size of the thermally-activated STZ also increases as compared to the strain-driven case.
  
This change in decay characteristics for thermally-activated surface as compared to bulk STZs can directly be tied to the geometry.  Specifically, for bulk STZs as shown in Figs. \ref{fig5}(a) and (b), as the strain-rate decreases and thermal activation plays a larger role, the nonaffine displacements decay more rapidly due to the fact that the deformation becomes more localized, i.e. fewer atoms move, but the magnitude of the motion increases due to the contribution of thermal energy.  In contrast, while the magnitude of the atomic motion also increases for the surface atoms that comprise the thermally-activated surface STZs, this also forces atoms near the surface to exhibit larger amplitude motions such that the zero stress state normal to the surface can be enforced.

\begin{figure} \begin{center}
\includegraphics[scale=0.25]{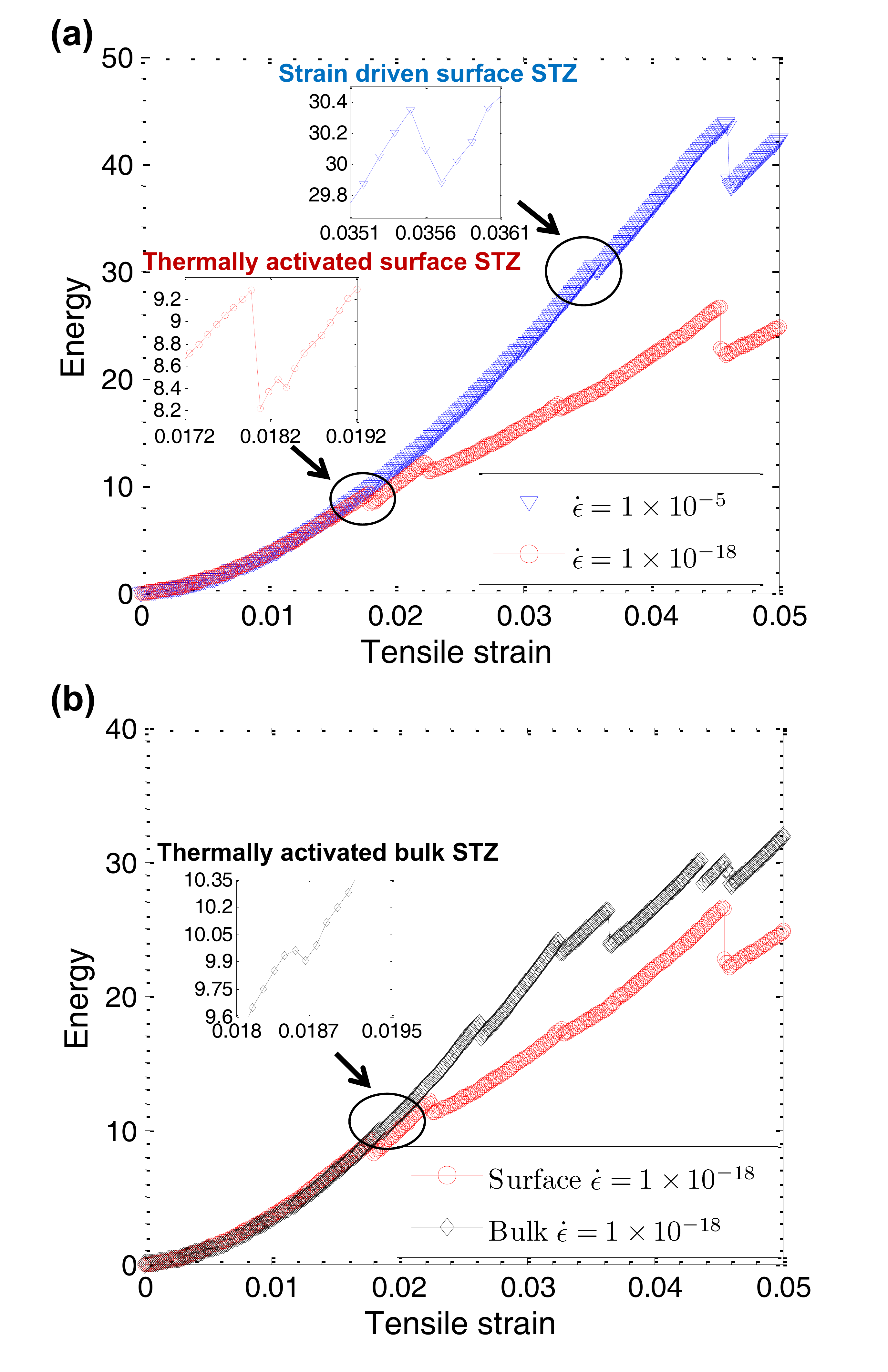} 
\caption{\label{drop}(Color online)  Representative potential energy versus tensile strain curves for (a) strain driven and thermally activated surface STZs and  (b) thermally activated bulk and surface STZs.}
\end{center} \end{figure}

This can also be quantified by calculating the average drop in potential energy corresponding to the formation of the first STZ, where a representative energy versus tensile strain curve for the surface STZ at high and slow strain rates is shown in Fig. \ref{drop}(a).  Before discussing the magnitude of the energy drop due to surface STZ formation, we note that the surface STZ nucleation strain is clearly smaller at the slower strain rate, which occurs because at the slower strain rate, the system has more time between loading increments to explore and cross over higher energy barriers on the PES, which implies that STZ nucleation can occur for lower tensile strains.

By calculating the average energy drop for both strain-driven and thermally activated surface STZs, we find that the average potential energy drop is 0.39$\epsilon_{SL}$ for the strain-driven surface STZs, while the thermally-activated surface STZs showed an average energy drop of 0.57$\epsilon_{SL}$.  On a normalized basis, the strain-driven surface STZs recover only about 1\% of the strain energy upon STZ nucleation, while the thermally-activated surface STZs recover about 7\% of the strain energy, where a representative comparison is shown in Fig. \ref{drop}(a).  Thus, the larger energy drop for the thermally-activated surface STZs results in a larger nonaffine displacement relaxation and slower nonaffine displacement field decay rate.  Finally, Fig. \ref{drop}(b) demonstrates that the energy drop is larger for thermally activated surface than bulk STZs, which is consistent with what is seen in Figs. \ref{fig5}(d) and \ref{fig7}(b), and is consistent with the notion of a more compliant surface as discussed above.  While we have not shown a figure doing so, this explanation comparing the energy drop can also be used to explain why for the bulk, as shown in Fig. \ref{fig6}(b), the strain-driven STZs decay slower than the thermally-activated ones.

We have already noted that 15 of 20 independent SLME simulations resulted in thermally activated surface STZs, whereas 11 of 20 independent SLME simulations resulted in thermally activated bulk STZs.  This suggests, perhaps not surprisingly, that the activation energy barrier for thermally-activated surface STZs is smaller than that required for bulk STZs.  

Finally, an important, and unresolved issue to consider in the future is the effect of the surface STZ type on the resulting failure, or shearbanding characteristics that occur in surface-dominated amorphous solids.  In particular, it is possible that the difference between thermally-activated and strain-driven surface STZs will lead to differences in the resulting shearband that forms, both in terms of the nucleation stress and strain, but also perhaps in terms of shearband orientation and size.

\section{Conclusion}

We have performed athermal, quasistatic atomistic calculations in conjunction with time-scale bridging atomistic calculations at both MD and experimentally-relevant strain rates to elucidate the structure and characteristics of surface STZs.  In the athermal, quasistatic limit which neglects temperature and strain rate, surface STZs exhibit similar decay rates to bulk STZs, though the direction of maximum nonaffine displacement is rotated away from the tensile loading direction.  Greater differences between bulk and surface STZs are found at room temperature and experimentally-relevant strain rates.  In particular, surface STZs exhibit a smaller decay rate of the nonaffine displacement field, and also show a greater tendency to exhibit thermally activated behavior that is only observed at experimental strain rates, in contrast to bulk STZs which show both thermally activated and strain driven behavior at experimental strain rates.  

All authors acknowledge the support of the NSF through Grant, No. CMMI-1234183. X.L. also acknowledges the support of Nenter \& Co., Inc., and NSF-XSEDE through Grant No. DMR-0900073.

\bibliography{biball}

\end{document}